\documentclass[manuscript,screen, nonacm, preprint,sigconf]{acmart}

\usepackage{xcolor}
\usepackage[draft,inline,nomargin,index]{fixme}
\usepackage{booktabs}
\usepackage{multirow}
\usepackage{tabularx}
\definecolor{ForestGreen}{RGB}{34,139,34}

\fxsetup{theme=color,mode=multiuser}
\usepackage{pifont}
\newcommand{\cmark}{\color{ForestGreen}\vspace{-2.3mm}\ding{51}}%
\newcommand{\xmark}{\color{red}\vspace{-2.3mm}\ding{55}}%

\AtBeginDocument{%
  \providecommand\BibTeX{{%
    \normalfont B\kern-0.5em{\scshape i\kern-0.25em b}\kern-0.8em\TeX}}}

\setcopyright{acmcopyright}
\copyrightyear{2023}
\acmYear{2023}
\acmDOI{XXXXXXX.XXXXXXX}

\acmConference[EAAMO '23]{ACM Conference on Equity and Access in Algorithms, Mechanisms, and Optimization}{October 30--November 1,
  2023}{Boston, MA}
\acmBooktitle{Boston '23: ACM Conference on Equity and Access in Algorithms, Mechanisms, and Optimization, October 30--November 1, 2023, Boston, MA} 
\acmPrice{15.00}
\acmISBN{978-1-4503-XXXX-X/18/06}

\begin{document}

\title[Dual Governance]{Dual Governance: The intersection of centralized regulation and crowdsourced safety mechanisms for Generative AI}


\author{Avijit Ghosh}
\affiliation{%
  \institution{AdeptID and Northeastern University}
  \country{USA}
  }
\email{ghosh.a@northeastern.edu}

\author{Dhanya Lakshmi}
\affiliation{%
  \institution{Peloton Interactive and Cornell Tech}
  \country{USA}
  }
\email{dl998@cornell.edu}
\renewcommand{\shortauthors}{Ghosh, et al.}

\begin{abstract}
  Generative Artificial Intelligence (AI) has seen mainstream adoption lately, especially in the form of consumer-facing, open-ended, text and image generating models. However, the use of such systems raises significant ethical and safety concerns, including privacy violations, misinformation and intellectual property theft. The potential for generative AI to displace human creativity and livelihoods has also been under intense scrutiny. To mitigate these risks, there is an urgent need of policies and regulations responsible and ethical development in the field of generative AI. Existing and proposed centralized regulations by governments to rein in AI face criticisms such as not having sufficient clarity or uniformity, lack of interoperability across lines of jurisdictions, restricting innovation, and hindering free market competition. Decentralized protections via crowdsourced safety tools and mechanisms are a potential alternative. However, they have clear deficiencies in terms of lack of adequacy of oversight and difficulty of enforcement of ethical and safety standards, and are thus not enough by themselves as a regulation mechanism. We propose a marriage of these two strategies via a framework we call Dual Governance. This framework proposes a cooperative synergy between centralized government regulations in a U.S. specific context and safety mechanisms developed by the community to protect stakeholders from the harms of generative AI. By implementing the Dual Governance framework, we posit that innovation and creativity can be promoted while ensuring safe and ethical deployment of generative AI.
\end{abstract}






\maketitle

\section{Introduction}

% \begin{figure}[htb!]
% \includegraphics[width=0.4\textwidth]{images/AIregulationmeme.jpg}
% \caption{A meme about the current lack of strict regulations in the AI landscape \protect\footnotemark}
% \centering
% \end{figure}

% \footnotetext{\url{https://staging.bsky.app/profile/deepfates.com.deepfates.com.deepfates.com.deepfates.com.deepfates.com/post/3juh2y3cstf2m}}

Generative Artificial intelligence (AI) has emerged as a fast-evolving subfield of Machine Learning (ML) that focuses on models that generate open ended content, such as text \cite{Introduc91:online}, code \cite{GitHubCo22:online}, images \cite{StableDi92:online, Midjourn72:online, DALL·E25:online}, videos \cite{RunwayEv83:online}, and even music \cite{AIMusicG54:online}. This technology has already started impacting many industries including education, entertainment, politics, and healthcare. However, as with any dual-use technology, there are significant ethical and safety concerns surrounding its use.

While the ability to create human-like content can be a powerful tool for creative expression, with proponents claiming that it democratizes creativity \cite{TheDawno26:online}, it also raises concerns about the potential for misuse, including the creation of misinformation, propaganda, and deepfakes. Images in a tweet that were generated using AI by Amnesty International \cite{AmnestyI98:online} illustrate a real-life harm of this technology due to misrepresentation of information. Amnesty International's Norway account artificially generated three images depicting protesters in a violent clash with law enforcement, stating that they did so to safeguard people on the ground. However, blurring the lines between truth and fiction sets a dangerous precedent, undermining work done to capture human rights violations by advocates. Additionally, there are concerns about the potential for generative AI to cause social harms, such as hallucinations \cite{bang2023multitask}, unfair bias \cite{naik2023social}, emotional manipulation \cite{Microsof38:online}, or encouraging self-harm \cite{HeWouldS82:online}.

On a more human note, people have argued that unbridled use of generative AI may eventually threaten to displace actual humans from the creative process \cite{AIiscopy61:online}, by decimating the livelihoods of artists, journalists, writers, musicians and other creatives.  Generative AI creators are already facing copyright battles \cite{Generati88:online} and liability issues in courts since these models are trained on work indiscriminately scraped from the internet and have the capability to copy the styles of individual content creators \cite{ghosh2022can}.

% It is critical that we begin considering policies and regulations that can help mitigate these risks and ensure that the development and deployment of generative AI is done in a responsible manner.
Countries have begun the critical work of drafting policy and regulation to mitigate these risks and promote the responsible development of generative AI, with the EU's AI Act describing protections and checks for a variety of AI systems, and China releasing a draft law to target generative AI and misinformation. With priorities and approaches varying by region, it is not surprising that the regulations also differ. For instance, the GDPR\footnote{\url{https://gdpr-info.eu/}} offers protections to a person based in Europe that a person in the U.S. does not have. Both proposed frameworks and the execution of existing standards are often incompatible and exhibit significant variance. 

Centralized regulations have a few other shortcomings. Regulations are often not technically specific \cite{masIntroducesFEAT}, with the vagueness then creating non-uniform interpretations across industry players. Hidden expenses in the form of legal and technical compliance teams \cite{europaImpactAssessment} can skew the competition unfairly towards incumbent companies, since small companies might not have the resources to properly navigate technical compliance. In the U.S., regulation moves at a glacial pace due to the nature of the democratic process with copious stakeholder input. By the time a law has passed, however, technology has often moved forward, with regulators playing catch up \cite{Whydoesi35:online}.

In the vacuum created by slow moving regulation, there is a growing community of researchers and developers who build tools and mechanisms for people to protect themselves from the harms of generative AI. For example, tools such as glaze \cite{shan2023glaze} (to protect artwork from being trained on), to erasure \cite{gandikota2023erasing} (to erase concepts from stable diffusion models), to Large Language Model (LLM) watermarking \cite{kirchenbauer2023watermark} (to detect whether a piece of text was generated by an LLM). These tools are distributed and can therefore be used by individuals to protect themselves. But this raises the question: can open source tools that aid in decentralized protection of stakeholders---or rather, subjects of AI harms---also aid in regulatory efforts? Is the more pragmatic solution a combination of both top-down and bottom-up approaches to fight AI harm? 

In this paper, we propose a middle ground -- a framework that we call \textit{Dual Governance.} This framework outlines a cooperative synergy between centralized regulations and crowdsourced safety mechanisms to protect stakeholders from the harms of generative AI in the U.S. It includes centralized regulatory efforts via U.S. federal agencies such as the FTC and NIST who set broad ethical and safety standards for generative AI, as well as a crowdsourced safety ecosystem comprised of of developers, researchers, and other stakeholders to audit generative AI models and develop tools that provide protections at the individual level. By striking this balance, we posit that the Dual Governance framework can promote innovation and creativity, ensure that generative AI is developed and deployed in a responsible and ethical manner, and protect stakeholders as technology evolves.

\section{Background}
\subsection{Harms of Generative AI} \label{sec:harms}

\begin{figure}[h]
\includegraphics[width=0.4\textwidth]{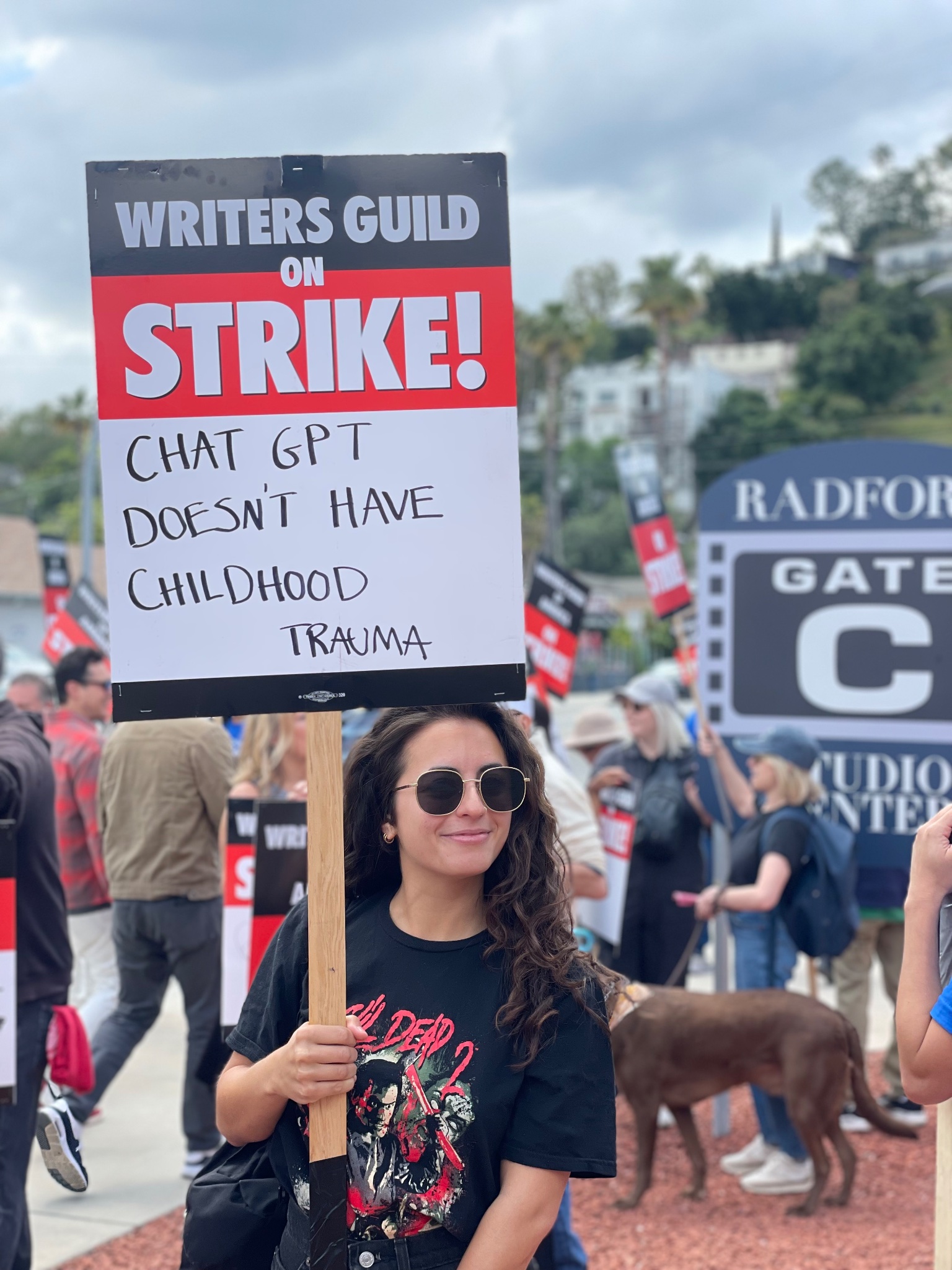}
\caption{A protester holding up a sign referencing ChatGPT at the 2023 WGA Strike\protect\footnotemark}
\centering
\end{figure}

\footnotetext{\url{https://twitter.com/fuckyouiquit/status/1654486969492054016}}

While certainly a powerful new paradigm in the ML landscape, Generative AI models bring with them pressing societal harms, that we discuss below.

One significant concern is the potential for Generative AI to spread misinformation. Because these algorithms can generate realistic-looking content, including text, images, and videos, they could be used to create fake news stories, social media posts, and even deepfakes. In recent news, the tool MidJourney AI \cite{Midjourn72:online} was used to generate fake images of President Donald Trump being arrested in New York, or Pope Francis in a puffer jacket, prompting the Pope to urge people to use AI ethically \cite{vaticannews}. Misuse of these technologies to spread false information has serious potential implications for the public's trust in information and could contribute to political instability or social unrest. Text generation models especially are particularly dangerous as they can seamlessly generate text that appears to be factual in context when they aren't \cite{bang2023multitask}, thereby exposing a naive user to potentially dangerous outcomes when used in high stakes use cases like healthcare \cite{ChatGPTi80:online}, or news reporting \cite{thevergeCNETPauses}.

Another concern is copyright abuse. For example, someone could use a Generative AI algorithm to create a piece of music or artwork that is very similar to an existing work, potentially leading to copyright infringement. This could has serious implications for artists and creators who rely on copyright protection to make a living. Furthermore, AI generated pieces of text and art in the style of living artists effectively takes their livelihood away from them, and further diminishes the incentives that artists and writers have in learning skills and creating new content \cite{ghosh2022can}. The threat of removing the human ingenuity component from content creation, and therefore livelihoods, has led to landmark lawsuits such as the ones by artists against Stability AI and Midjourney \cite{stablediffusionlitigationStableDiffusion}, and community protests, like the Writer's Guild of America Strike \cite{timeWritersStrike}.

Intertwined with copyright abuse are privacy and consent issues. Generative AI models are trained with text and images indiscriminately crawled from the internet, with little regards to whether it is personal information, copyrighted work, or harmful content \cite{birhane2021multimodal,ghosh2022can}. This data can then be used for truly malicious outcomes like generating deepfakes, or new content in the style of a particular person without their consent. Models have also been repeatedly shown to often memorize verbatim and easily regurgitate information in training data, sometimes private data like credit card information or addresses \cite{carlini2021extracting, carlini2023extracting}, and have also been shown to be vulnerable to prompt injection attacks \cite{greshake2023more}. 

Finally, there is a potential for Generative AI to reinforce existing biases and inequalities. Studies have shown that biases exist in the outputs of both text generation models \cite{zhuo2023exploring} and image generation models \cite{naik2023social}. As more and more synthetically generated content is released into the world and on the internet, biased content has a worryingly large capacity to spread racist, sexist, ableist, or other kinds of socially inappropriate content at a scale that publishers or moderators currently have no capacity to deal with \cite{Scifipub34:online}.

% Allude to Neurips paper etc

% [some leads for research]
% - misinformation use cases: Bing + OpenAI, CNET using GPT to write financial articles, DoNotPay - the company trying to offer dubious advice
% - using this kind of tech to generate marketing campaigns, drafts for legal or medical documents can also be super harmful 

\subsection{Existing Governance Models}\label{sec:bgGovernance}
This section briefly covers the regulations introduced by various countries and proposed policies across academic papers, as well as tools and methodologies to assess models for technical risks. 

\paragraph{\textbf{USA:}} The proposed Blueprint for an AI Bill of Rights by the White House Office of Science and Tech Policy \cite{whitehouseBlueprintBill} identifies five principles around safety, discrimination, privacy, and human considerations to be used to guide the development of AI systems, along with details on actualizing these principles in model development life cycles. Additionally, bills like the ASSESS AI Act \cite{S135611824:online} represent ongoing legislative efforts to mandate the responsible deployment of AI.

\paragraph{\textbf{EU:}} The AI Act takes into consideration data quality, privacy, and ethics concerns that arise from how AI systems are used. It categorizes these systems into four levels of risks, unacceptable, high, limited, and minimal, according to which there are either restrictions on the usage of AI (such as with real-time surveillance devices), or requirements for compliance audits. The EU has also taken strides to address regulations for generative AI by considering a tiered approach that will subject models to different levels of compliance based on context and level of risk. For example, the requirements that apply to foundational models will be different from those that apply to the fine-tuned models that are used for specific purposes, thereby ensuring that not all models will have to conform to the strictest requirements. The first public case of a European regulator taking an action against a generative AI model was in March 2023, when Italy's data regulator, GPDP, banned ChatGPT in Italy due to privacy and age restriction concerns \cite{ChatGPTb19:online}. OpenAI was eventually forced to technically comply with these requirements and add a training opt out feature before Italy lifted the ban \cite{ChatGPTr45:online}.

\paragraph{\textbf{Singapore:}} The government has released a set of Fairness, Ethics, Accountability, and Transparency (FEAT) principles \cite{masIntroducesFEAT} for monetary organizations  that outline principles that should be considered in building these systems. An MVP tool created by the government, AI Verify, provides a framework to test a subset of models against eight specified AI principles and generate a report that outlines how well the model performed.

\paragraph{\textbf{China:}} China has created specific regulations for recommendation algorithms that include mandatorily registering recommendation algorithms  with the internet regulator and submitting them to the Cyberspace Administration of China (CAC) for a security assessment \cite{insideprivacyChinaProposes}. For generative AI, the CAC released a draft of rules that focus on content moderation and misinformation that also mandate a security assessment by the CAC’s providers before launch. However, there are some rules in this draft that may be hard to enforce due to the generational nature of these systems, such as the content moderation policy: “Ensure that content created by generative AI is true, accurate, and free of fraudulent information;”. What this policy means by “true” will need a lot of clarification, and methods to test and metrics to validate this will also need to be defined by regulators or authorized third parties. 

\paragraph{\textbf{Proposed policy papers:}} There are a number of policy papers in the literature offering critiques, disagreements and proposals. They range from articles arguing that quick regulation amidst what is described as a ``tech panic'' would do more harm than good \cite{datainnovationTechPanics}, to papers that propose requirements in addition to what is specified by the EU specifically for general purpose AI including transparency requirements and mandatory but limited risk assessments \cite{hacker2023regulating}. Additional downstream risk assessments of generative models either by AI providers \cite{policyreviewChatGPT} or by the middlemen using it in products \cite{hacker2023regulating} have been proposed so as to better assess context-specific harms of a general-purpose AI system. Another article \cite{lawfareblogDevelopingRegulation} proposes incorporating recourse for users so that individual rights are not ignored, and creating an ``FDA for algorithms'' so that only licensed algorithms are in use, the latter of which is similar to China's CAC proposal. Finally, in this \cite{springerSafeReliance} article by Grandeur et al., a bottom-up approach is suggested as a method of regulation of AI, with the focus being on self-regulation by consumers, developers, academia, and companies, and minimal, supplemental regulation of AI by a separate government agency. This includes collective agreement on a set of values, transparency, and respecting implicit guidelines set in the industry. We submit that while this is an ideal scenario, collaboration between governments and the community is a far more effective way of maintaining accountability.   

\paragraph{\textbf{Tools and Risk Frameworks:}} To supplement regulation, risk assessment frameworks and tools have become an effective method at identifying and mitigating AI risk. Various academic scholars, companies, and independent organizations have proposed methods to assess harms at various stages of a model’s life cycle, some even by drawing upon existing processes in different industries. For example, Rismani et. al. \cite{rismani2023plane} explore the adaptation of System Theoretic Process Analysis (STPA), a safety engineering framework, for ML risk assessment and mitigation. The algorithmic auditing framework described in this paper by Raji et al. \cite{raji2020closing} provides ML practitioners with a method to test their models for harms at different stages of the development process. MITRE has also released ATLAS (Adversarial Threat Landscape for Artificial-Intelligence Systems), a knowledge base of techniques and tactics that describe ways that models are susceptible to attacks in the real-world so as to better understand the scope and impact of the identified harms. 

\paragraph{\textbf{A note on copyright:}} Since Generative AI models train on a large amount of data scraped off the internet to generate text, images, and audio, additional regulations around copyright infringement apply to these systems. Existing copyright laws around the world are mainly focused on art made by humans, with originality of the art being a large part of the laws. For AI-generated art, however, attempts to allow copyright protection for AI-generated art have generally been divisive. While Creative Commons has argued that using AI for art should be permitted under the Fair Use law \cite{creativecommonsFairUse}, the U.S. Copyright Office recently changed a decision \cite{harvardZaryaDawn} removing copyright protection from images in a graphic novel, \textit{Zarya of the Dawn} that were generated using Midjourney while maintaining protections for the original work (text, selecting the art) in the novel. While these decisions are being made in a world where AI image generation tools are scraping images from the internet indiscriminately, we may need to change our approach as the data used for these tools becomes more regulated, and more artists begin using image generation AI as tools to aid them in making new art. There are already community-sourced initiatives like Responsible AI Licences (RAIL) \cite{contractor2022behavioral} aiming to create a middle ground. An ideal solution would provide protections to the artists who make the final artwork, as well as those whose art has been trained on by the AI tool. While it may not be possible to reach this ideal state, moving copyright laws in that direction might be the first step. 

\section{Policy Scope}\label{sec:scope}

In this section, we attempt to carefully delineate the scope of the policy framework we propose. We identify the stakeholders who will be affected by the policy, identify the values that will guide the development of the policy, identify the domains that the policy will impact, identify the potential harms that the policy may cause, and finally prioritize the issues and goals that the policy addresses.

\paragraph{\textbf{Domains:}} Our policy suggestion exclusively aims to cover open ended commercial generative models (systems that generate text, images, video, and audio).

\paragraph{\textbf{Stakeholders:}} Our policy suggestion aims to cover the people using generative AI products, meaning the set of all consumers of generative AI products, government regulators, and rational commercial agents (big and small companies that sell generative AI products). 

\paragraph{\textbf{Harms:}} The harms that the policy suggestion aims to cover are copyright issues, misinformation, economic impacts, consent issues, and societal biases. A detailed discussion of these harms are in \ref{sec:harms}.

\paragraph{\textbf{Values:}} The values that guide the development of the policy are safety, innovation, and agency. Our proposed framework aims to keep people using generative models safe, while also providing agency to individuals in the form of copyright and privacy protections, and actionable recourse. While keeping these values intact, we also aim to promote technical innovation in the generative AI landscape.  

\subsection{Criteria for an Effective Governance Framework} 

In building an integrated framework, we would like to focus on a few key objectives. Firstly, through applying this framework, generative AI should not cause the harms described in \ref{sec:harms}. Secondly, we want to build an evolving framework that is mindful of the technical feasibility of the compliance requirements. Thirdly, we do not want to hinder innovation or competition among companies of all sizes. Finally, we want the framework to provide access to communities at large without violating existing data privacy laws. 

To meet these objectives, we have constructed criteria that we believe should be considered while building an integrated framework, such as the Dual Governance Framework that we are proposing.  

\begin{enumerate}
    \item \textbf{\textit{Clarity:}} The framework should contain policies that are understandable and are associated with one or more technically feasible solutions. This is important in making the framework accessible to and easy-to-use for consumers of varying backgrounds. 
    \item \textbf{\textit{Uniformity:}} The technical specifications for compliance should be interpreted uniformly across the stakeholders who are authorized to implement it, along with potentially having templates for development.  
    \item \textbf{\textit{Availability:}} The framework and its cost of usage should be tailored so that smaller companies can use it to comply with regulations easily, and encourage innovation. The tools available through the framework must be generally applicable to many models. 
    \item \textbf{\textit{Nimbleness:}} Having \textit{nimbleness} as a criteria will allow the framework to provide consumer safety while centralized regulation is being debated and finessed by governments. Hence, the proposed framework must be able to adjust quickly to new developments in generative AI and make new tools available for consumers.
    \item \textbf{\textit{Actionable Recourse:}} To preserve agency, consumers should be able to contest decisions made by a system that uses AI for decision making and request an alternative, non-automated method of decision making. They should also be able to report suspected discrimination or violation of laws by a system they encounter.
    \item \textbf{\textit{Transparency:}} The tools and mechanisms that are a part of the framework should be public, where reasonable. This is advantageous for all stakeholders. For big tech companies, regulators will have more confidence about their internal implementations for similar requirements. For the consumers, this allows easy access to collaboration. Transparency can be achieved via regulatory requirements, however, it could lead to big tech companies establishing monopoly over the development of transparency tools, with smaller tech companies becoming limited in their choice. 
    
\end{enumerate} 

\begin{table*}[]
\centering
\begin{tabularx}{\textwidth}{@{} l l X X @{}}
\toprule
\textbf{} & \textbf{Parent agency} & \textbf{Purview} & \textbf{Policies implemented/examples of actions taken} \\
\midrule
\textbf{NIST} & Department of Commerce & Builds standards for AI and risk assessment methodologies & AI Risk Management Framework \\
\midrule
\multirow{3}{*}{\textbf{FTC}} & \multirow{3}{*}{Independent Agency} & Ensures that commercial products using AI do not harm consumers and investigates violation of AI laws against consumers across various industries & Complaint against Bronx Honda for discriminating against African-American and Hispanic car buyers \cite{ftcAutoDealership} \\
\cmidrule{3-4}
& & Implements rules to ensure that exaggerated claims are not made about AI products, provide information about what and how data is being collected, and ensure that fair decisions are being made about consumers &  Business guidance on using AI in products \cite{ftcKeepYour, ftcAimingTruth} \\
\midrule
\multirow{5}{*}{\textbf{EEOC}} & \multirow{5}{*}{Independent Agency} & Examines use of AI in employment decisions & Engaged in a public hearing to obtain comments about the use of automated systems in employment decisions \cite{eeocEEOCHearing} \\
\cmidrule{3-4}
 & & Develops AI and Algorithmic fairness initiatives and ensures AI tools do not violate the Americans with Disabilities Act (ADA) & Guidance for employers on how to comply with the ADA while using AI in employment processes \cite{eeocAmericansWith} \\
\cmidrule{3-4}
 & &  Evaluates workplaces to ensure that they are free of race and color discrimination & Lawsuit against iTutorGroup for age discrimination \cite{eeocEEOCSues} \\
\midrule
\multirow{3}{*}{\textbf{CFPB}} & \multirow{3}{*}{Federal Reserve} & Protects consumers from financial risk due to AI products, including protection against algorithmic marketing, complex AI-dependent technology, algorithmic bias in home valuations and banking & Circular to protect the public from black-box credit models using complex algorithms \cite{consumerfinanceCFPBActs} \\
\cmidrule{3-4}
 & & Allows tech workers to submit whistleblower complaints \cite{consumerfinanceCFPBCalls} against financial institutions & Interpretive rule on the responsibilities of digital marketers regarding algorithmic ads \newline A proposal for a registry to detect repeat offenders \cite{consumerfinanceCFPBProposes} \\
\midrule
\multirow{3}{*}{\textbf{CRT}} & \multirow{3}{*}{Department of Justice} & Enforces constitutional and statutory civil rights protections, including involvement in cases pertaining potentially biased AI systems & Statement of Interest in Statement of Interest in Fair Housing Act Case \cite{justiceJusticeDepartment} \\
\cmidrule{3-4}
 & & Releases strategy plans to help the DOJ adapt to the changing AI ethical landscape, and better support evolving AI requirements from various government bodies & Artificial Intelligence Strategy for the U.S. Department of Justice \cite{AIStrategyDOJ} \\
\bottomrule
\end{tabularx}
\caption{Overview of U.S. federal agencies and their roles in regulating AI}
\label{table:regulating-AI}
\end{table*}

\section{Centralized Regulation in the U.S. Context}\label{sec:centralizedregs}

There has been a surge of requests for AI regulation from multiple fronts in recent times. The opinion piece by Dr. Rumman Chowdhury \cite{wiredDesperatelyNeeds} makes the case for a global, independent oversight board for AI to safeguard the public from AI harms. Section \ref{sec:bgGovernance} describes efforts to regulate AI and specifically, generative AI as well across different countries. In this section, we would like to focus on regulatory developments in the U.S., and shed light on the capabilities of and actions taken by different agencies within the country. In particular, we will be looking at the Federal Trade Commission (FTC), the Department of Justice's Civil Rights Division (CRT), the Consumer Financial Protection Bureau (CFPB), the Equal Employment Opportunity Commission (EEOC), and the National Institute of Standards and Technology (NIST). These federal agencies play an important part in enforcing civil rights, non-discrimination, fair employment regulations, consumer protection, as well as implementing standards. While NIST is working on developing an AI Risk Management Framework \cite{nistRiskManagement} following the National Artificial Intelligence Innovation Act of 2020 \cite{congressAIBill}, the EEOC, FTC, CFPB, and CRT have also been taking action to protect consumers across different industries from risks of rapidly evolving AI systems, even announcing a partnership for joint enforcement against discrimination and bias in AI \cite{eeocJointStatement}.

Table 1 describes some of the policies and efforts taken by the above-mentioned federal agencies to regulate AI in the U.S. Policies dealing with AI harms across a variety of focus areas, including finance, employment, and law are covered by at least one these five agencies. AI regulations created in different domains ensure that consumer rights are protected across multiple avenues. For example, with direction from the Consumer Protection Act, the CFPB and other federal agencies worked to outline policies to prevent algorithmic bias in home valuations. On the legal front of this issue, the CRT filed a statement of interest in the Fair Housing Act's (FHA) use of automated systems to appraise homes, with the intent being to emphasize that tenant screening policies by automated systems still fall under the FHA. Other examples of focus on AI include the EEOC's Strategic Enforcement Plan (SEP) \cite{eeocArtificialIntelligence} with its intent to enforce non-discrimination laws on automated decision making systems, the CRT's inclusion of governing AI systems and aim to shape DOJ laws and approaches to AI in its strategy for 2020 and 2023-24 \cite{AIStrategyDOJ}, and the circulars regularly published by the CFPB and the FTC providing advice and warnings on specific usages of AI in consumer-facing products \cite{consumerfinanceCFPBActs, ftcKeepYour, ftcAimingTruth}.

These agencies have also indicated their interest in regulating generative AI. The EEOC has held public hearings on topics including generative AI to get feedback from the public. The DOJ and the FTC are analyzing generative AI tools for anti-competitive behavior \cite{thevergeGovernmentGearing}. On the other hand, the CFPB is monitoring the use of chat-GPT and similar generative AI tools by banks \cite{bankingdiveBanksChatGPTlike}. The FTC's commissioner, Alvaro M. Bedoya, in his prepared remarks about generative AI, reiterated the applicability of acts such as the Fair Credit Reporting Act and the Equal Credit Opportunity Act on AI used in today's society. He also advocated for transparency in models and emphasized the need for researchers, civil society and government to analyze and stress-test models \cite{ftcPreparedRemarks}.

The history of collaboration between these industries is another advantage. NIST released a special report describing a standard for investigating bias \cite{nistTowardsStandard} that referenced work done by CFPB and other agencies. The new joint initiative from the EEOC, FTC, CFPB, and CRT, will allow for more effective collaboration, and potentially the standardization of terminology, policies by acknowledging the existing overlap between them. Their indicated areas of interest include imbalanced datasets with historical bias, lack of context for the system in which AI is being used, and the black-box nature of large algorithms. These focus areas can also signal to companies what to focus on while building AI products. Building best practices and identifying harms in these areas during development could also lead to building ethics-forward workflows.

\subsection{Can centralized regulation be enforced effectively?}\label{sec:regcons}

While governmental regulations cover a wide range of use cases, and do help in setting a national standard, it is to be acknowledged that the non-specificity of these policies make them hard to enforce, and sometimes may be counterproductive as they allow companies to assert compliance without addressing technological harms. Defining audit methodologies, performing compliance audits, and identifying qualified agencies and third parties is also an extensive process. Coupled with the time it takes for regulations to catch up to this rapidly evolving space, this presents a significant downside to centralized regulation.

Governments are not immune to pressure from large technology companies, regardless of their expertise in the development and usage of AI. In the recent past, in the interest of integrating it into their own products for commercial gain, companies leaned on the EU to create exceptions for general purpose AI \cite{techcrunchReportDetails}. Even if the EU ignores these persuasions, these companies may yet succeed in other countries. While it remains to be seen how much these tech companies affect regulation on generative AI, an ideal regulatory framework should be impartial to such influences.  

The increase in calls for regulation has also been met with resistance from the industry, with a common complaint being that ``unnecessary regulation'' \cite{techcrunchLawmakersTiered} will hinder innovation and cost companies a lot of money. While a portion of these complaints are from tech companies trying to get ahead of their competition quickly, that this cost of compliance will impact smaller and emergent startups. Research from the EU provides additional perspective: estimating the cost of compliance to be between 4 to 5\% of investment in high-risk AI by 2025 \cite{europaImpactAssessment}. The mandate for compliance verification could be offset by the increase in research of verification methodologies, which in turn will lead to better tools and frameworks in this space. For the policy makers, these concerns should be taken as more incentive to work towards figuring out the right balance that protects users while encouraging the development of General Purpose AI (GPAI) models through an iterative process. Not doing so will lead to bad regulation, which will result in additional work not just for technologists, but also for policy enforcers. For example, the UK’s proposal \cite{computerworldGovernmentsStrategy} to allow downstream regulators in each industry handle tackling AI harms for their own use cases understandably was met with resistance, as it would have led to more work for regulators who did not have the same levels of AI knowledge as well as inconsistencies in terminology and policies across industries.

Enforcing policies that cover a large set of harms, ensuring that they are not ambiguous or too specific, that can also be audited in a reasonable amount of time, and that can keep up to innovation in this field is a complicated challenge, and is one reason people are looking at crowdsourced tools as an intermediary alternative. 

\section{Crowdsourced Safety Mechanisms}

So far, we have focused on the current and proposed landscape of regulations in the space of generative AI. Generative AI is a rapidly evolving technology, and regulators may struggle to keep up with the pace of innovation and the constantly changing landscape of the field. Additionally, regulations may not be able to account for all of the potential edge cases of harms of generative AI, which can be diverse and complex. Missing from the rightful calls for national regulators to step in before the harms get worse is the practicality of complying to those calls, as we discuss in \ref{sec:regcons}. The open source communities and the academic research communities have in addition, started looking at technical ways to counter the harm posed by generative AI. We discuss some of these techniques below.

\paragraph{\textbf{Prevent Unwanted Training:}} \citet{shan2023glaze} have developed an technique (and app) called Glaze that adds almost imperceptible perturbations to artwork to interfere with AI models' ability to read data on artistic style, making it harder for generative AI technology to mimic the style of the artwork and its artist. The app helps artists fight back against data scrapers' incursions and at least disrupts their ability to rip off hard-worked artistic style without them needing to give up on publicly showcasing their work online. The app aims to equip artists with a free tool to defend their work and creativity from being indiscriminately ingested by image generating AIs.

\paragraph{\textbf{Watermarking:}} Text outputs generated by state-of-the-art LLMs, such as ChatGPT, are so convincingly human-like that there are concerns that these models can be used for plagiarism. Educators are especially concerned that students might use ChatGPT to write essays or code, thereby finding a way to cheat in their exams. Work by \citet{kirchenbauer2023watermark} shows a technique to watermark the outputs of LLMs so that they continue to seem human-like, but can be easily detected as a LLM output as opposed to human created text.

\paragraph{\textbf{Editing Trained Models:}} An interesting new line of research has started to look at how trained models can be edited to achieve certain properties. There are methods to erase specific concepts from trained text to image generation models \cite{gandikota2023erasing}, and methods to edit or delete memories or facts from a transformer based large language model \cite{meng2022mass,meng2022locating}. These techniques serve as secondary recourse after problematic or proprietary data has been used to train generative AI models.

\paragraph{\textbf{Deepfake prevention:}} Fake images and videos of real people are an ever growing misinformation threat that can potentially cause political or communal unrest. \citet{yang2021defending} have proposed a method to defend against deepfakes by adversarially modifying images of the faces of potential victims before uploading them on the internet, so that deepfake models generate undesirable artifacts when trained on these adversarial faces and can easily be spotted as fake.

\paragraph{\textbf{Data Provenance Tools:}} Community efforts to carefully document the massive, internet-size training datasets of generative AI models have sprung up in response to the rampant scraping of data by AI companies. There are tools\footnote{\url{https://rom1504.github.io/clip-retrieval/}} allowing artists and people to see if their images appeared in LAION-5B \cite{schuhmann2022laion}, the training dataset used to train DALL.E \cite{DALL·E25:online}, Midjourney \cite{Midjourn72:online} and Stable Diffusion \cite{StableDi92:online}, and similar tools\footnote{\url{https://www.washingtonpost.com/technology/interactive/2023/ai-chatbot-learning/}} to search within the massive text datasets scraped from the internet that LLMs like ChatGPT are trained on.

\paragraph{\textbf{Responsible AI Licensing:}} RAIL (Responsible AI Licenses) \cite{contractor2022behavioral} is a new community standard that provide developers with the ability to limit the use of their AI technology to prevent its application in irresponsible and harmful ways. These licenses contain clauses regarding the usage of AI that grant permission for specific use-cases while restricting certain other use-cases. If a RAIL license allows derivative works, it also necessitates that any downstream derivatives, including modification, redistribution, and repackaging of the licensed AI, must adhere to the behavioral-use restrictions outlined in the license. Notably, stable diffusion is open sourced with a RAIL license\footnote{\url{https://huggingface.co/spaces/CompVis/stable-diffusion-license}}.

\paragraph{\textbf{Bug Bounties and Hackathons:}} Red teaming exercises by the community are a valuable tool in the fight against the unchecked harms of generative AI, realized by bug and bias bounties \cite{Abiasbou88:online}, and via hackathons \cite{AIVillag22:online}. These spaces are usually inhabited by people who are independent tech enthusiasts and usually not part of either regulatory bodies or the tech companies being red teamed.

\subsection{Are crowdsourced technical protections enough?}

While crowdsourced tools and mechanisms to protect consumers from the unchecked harms of generative AI have several advantages over government regulations only, namely open and clear technical specifications, rapid evolution with technology, and better public trust in the defense mechanisms due to transparency -- they have a major caveat: enforceability. Only depending on community tools without any government intervention is, unfortunately, a form of guerrilla defense framework that can quickly descend into chaos. At best, these are stopgap measures, but at worst, it sends an incorrect priority messaging: in the face of the rapid and well documented harms of generative AI models, is every person on their own? Moreover, accountability becomes a major concern: if there are bad actors that a crowdsourced tool has identified, who is going to administer punishment? Ultimately, for better or worse, centralized regulatory agencies still serve important executive roles that the decentralized community of stakeholders cannot perform on their own. Both top-down and bottom-up approaches are required for a robust, dynamic framework against the threats of generative AI.

\section{The Dual Governance Framework}
In the previous sections, we have enumerated a variety of centralized regulations and crowdsourced safety mechanisms, and examined the U.S. federal agency landscape in detail. We also have defined the scope, stakeholders and criteria for the framework. Now, we will describe the Dual Governance Framework in detail, and provide a comparison against other frameworks.

\subsection{Overview}\label{sec:DualGovOverview}

Our proposed dual governance mechanism seeks to combines both top-down, centralized government regulation and bottom-up, decentralized community safety efforts -- by creating a regulatory framework that sets minimum standards for AI systems and requires compliance with those standards, while also providing opportunities for stakeholders such as users and experts to contribute to the ongoing assessment methodologies and improve AI systems. In doing so, this framework balances the benefits of centralized regulation with the advantages of crowdsourced safety mechanisms.

At a high level, the framework would involve an existing federal regulatory body (or a collaboration between agencies), that would establish guidelines and standards for the development and use of AI systems. We do not explicitly propose the creation of a new regulatory body, as policy and legal experts have shown that there are ways for existing agencies and patchwork of regulatory bodies to be flexible enough to govern AI \cite{TheBestW59:online}. We also have a potential good example in the united approach introduced by the U.S. federal agencies in \ref{sec:centralizedregs} combining regulatory bodies to govern AI. Guidelines issued by the agency or collective would set minimum standards to limit the harms and propagate the values we discuss in \ref{sec:scope}. Companies and organizations would be required to comply with these guidelines, and the centralized agency or collective would have the power to enforce compliance through inspections, fines, and other measures.

In addition to centralized regulation, the framework would also incorporate crowdsourced safety mechanisms, which would involve stakeholders in the ongoing assessment and improvement of AI systems, with the aim of making  compliance easier. This could take several forms, including:

\begin{itemize}
    \item \textbf{Public feedback, town halls, and reporting:} The government mandates companies as well as agencies to provide a mechanism for users to report issues and provide feedback on AI systems. Town halls are also conducted to gather direct feedback and suggestions from consumers. This feedback could be used to identify issues and areas for improvement.

    \item \textbf{Providing alternatives:} Government agencies provide a way for consumers to take action when they believe they have been subject to incorrect or unfair decisions from AI systems. This could range from defining processes for requesting the decision to be reviewed by humans, to filing lawsuits in situations of significant damage. 

    \item \textbf{Expert review:} Federal agencies engage experts in the field to review AI systems and provide feedback on their safety and efficacy. This could include academics, industry experts, and other stakeholders.

    \item \textbf{Community audits and research:} Community auditors review AI systems and provide feedback on their biases, safety and efficacy via hackathons and bounties. Parallely, the research community publishes defense mechanisms for decentralized protections. The federal agencies continue to assess the outcomes of these community efforts and issue up to date best practices, effectively informing future regulation.
\end{itemize}

Overall, the dual governance mechanism would aim to strike a balance between centralized regulation and crowdsourced safety mechanisms, with the regulatory body setting minimum standards and enforcing compliance, while stakeholders provide ongoing feedback to improve the safety and efficacy of AI systems. We discuss these two aspects in further detail in \ref{sec:eval} and \ref{sec:new}.

\subsection{A path for evaluating crowdsourced safety mechanisms} \label{sec:eval}
Incorporating crowdsourced mechanisms is a cornerstone of our framework. Simply creating regulations for generative AI systems could lead to lending legitimacy to potentially dangerous technology that could spread misinformation or use people's data in malicious ways. Crowdsourced mechanisms will grant consumers autonomy to identify and choose how their personal data is being used and provide alternative paths to recourse. Centralized regulatory agencies can also monitor how policies are being interpreted, validate them, or provide feedback. Champions of open-source methodologies can not only implement tools that satisfy central regulations, but also technically implement and shape regulations by voicing opinions and sharing solutions. Deputizing open-source tools will enable consumers and tech companies alike to trust in the tools and utilize them to comply with regulations. We define the steps that the evaluation process should contain:

\begin{itemize}
    \item \textbf{Who does it? }Identifying government agencies like NIST or the FTC to dedicate resources to processing new crowdsourced mechanisms. Since these agencies already work on setting policies and risk management frameworks, the lift required to validate new mechanisms is very small. Alternatively, these agencies could authorize third-party companies to process these mechanisms, while also committing to routinely audit these companies.
    \item \textbf{When does it happen? }Defining a timeframe in which these new mechanisms will be processed. This could take many forms, such as directing an agency like NIST to go through and certify new mechanisms every six months. The agencies could also be given authority to decide when a new mechanism needs full congressional approval. 
    \item \textbf{How are mechanisms certified? }Creating a transparent set of requirements and tests to verify these mechanisms. The requirements should include testing the mechanisms for bias, validating that it meets its stated objectives, and ensuring that the tool is public. Technologically, this could take the form of a GitHub pull request or a JIRA ticket. The agencies could also rely on consumer reports of how the tool works, provided evidence is shared to support their claims. Over time, a test methodology could be developed that runs a number of tests on the proposed tool.
    \item \textbf{How does certification work? }Authorizing or certifying the tool and adding it to a registry. Having a centralized place where stakeholders can access tools that have been validated by these agencies would allow them to be widely used and tested against different contexts. There should also be a time limit after which this certification expires, to ensure that the tools are up-to-date with the latest standards. 
\end{itemize}

\subsection{Adding and creating new regulations} \label{sec:new}
As mentioned in \ref{sec:regcons}, the time taken to regulate innovations in AI is a significant challenge. In the U.S., passing a law requires consensus from the House, the Senate and the President, and depending on the political scenario and the policy priorities of the nation, getting new regulations passed can potentially take years. To maintain \textit{nimbleness}, policies regarding new AI systems need to be integrated so federal authorities can remain in-step with developments in the field. To do so, we propose a review of and research into papers and tools about new developments in AI, with authority being given to an agency like NIST so that they may discern valuable tools and papers from conferences like FAccT, NeurIPS, etc., and make them available to use. Agencies can use their existing infrastructure to regulate AI in many ways, a point that has been argued by Lina Khan, the chairwoman of the FTC \cite{nytimesOpinionLina}. The growing body of technologists employed by the FTC, CFPB, etc., can allow them to draw on knowledge from the Free and Open Source Software (FOSS) community to create apply existing laws to AI systems. A review of new regulations and policies every few years would help propagate potential changes, like de-commissioning existing tools or policies or creating new ones. This is a tedious but necessary process, and ultimately will help in keeping this framework accurate and timely.

\subsection{Satisfying the prescribed criteria}
The Dual Governance framework aims to bring \textit{clarity} to existing regulations by associating them with technical interventions. Having a registry of existing mechanisms that provides detailed information on the tools and establishing that regulations are interpreted the way that centralized agencies intend satisfies \textit{uniformity} and \textit{transparency}. \textit{Availability} is ensured by making tools available to consumers in a cost-effective and accessible manner. Requiring a review of new frameworks and existing tools periodically, while tedious, allows this framework to change with the times, therefore fulfilling the \textit{nimbleness} criterion. Finally, centralized regulatory agencies like CFPB \cite{consumerfinanceCFPBCalls} already have methods to allow whistleblowers to alert the agency to malpractice. Taking inspiration from this and the policy for human alternatives proposed in the Blueprint for an AI Bill of Rights\cite{whitehouseHumanAlternatives}, the framework defines methods for alternative action in \ref{sec:DualGovOverview}. Table 2 compares the Dual Goverance framework with centralized regulation and crowdsourced safety.   

\begin{table*}[h]
\centering
\begin{tabularx}{\textwidth}{@{} l *{3}{>{\centering\arraybackslash}X} @{}}
\toprule
\textbf{Criterion} & \textbf{Centralized Regulation} & \textbf{Crowdsourced Safety} & \textbf{Dual Governance} \\
\midrule
Clarity &\xmark&\cmark&\cmark\\
Uniformity &\cmark&\xmark&\cmark\\
Availability &\xmark&\cmark&\cmark\\
Nimbleness &\xmark&\cmark&\cmark\\
Actionable Recourse &\cmark&\xmark&\cmark\\
Transparency &\xmark&\cmark&\cmark\\
\bottomrule
\end{tabularx}
\caption{Comparison of Governance Models with Dual Governance}
\label{table:all-governance-models}
\end{table*}

% - Risk Assessment Frameworks/ Auditing/ In depth assessments
% - If risk assessment and mitigation tools are open source, then open source frameworks for AI/ML will also be able to easily incorporate them, instead of leaving it up to practitioners 
%  Open source tools will allow people to have some autonomy over how their information is used, and provide them with an intermediate path to recourse while regulation enforcement catches up.   

% Need to talk about the limitations of this paper
\section{Limitations}

There are some key limitations to our work. The main limitation has to do with the scope of the problem that we have set out to tackle. While consumer facing, open ended text and visual media generation models are certainly the most visible use cases of generative AI, there are several other use cases that we do not cover here, such as when generative AI models are packaged inside other products, such as office suites \cite{Announci57:online} and healthcare \cite{TruvetaT17:online}, because these use cases are intrinsically linked with different sets of harms and values, and consequently different regulatory agencies, that we do not cover. Our framework is, by design, U.S. specific and domain specific, and therefore necessarily incomplete.

Furthermore, an important consideration in our framework design is the assumption that the stakeholders responsible for complying with the framework are rational commercial agents, and therefore it can be reasonably expected that they will comply with government mandates, copyrights and licenses. Our design deliberately does not include bad actors -- for instance people who knowingly violate\footnote{\url{https://www.reddit.com/r/StableDiffusion/comments/wv2nw0/tutorial_how_to_remove_the_safety_filter_in_5/}} licenses attached to open source generative AI models. Open sourcing a model crosses into the domain of liability issues, and governing them or enforcing rules on these bad actors is an important but completely different set of challenges that we do not address within our framework. Additionally, the rapid development of generative AI has given rise to existential harm discussion from several members of the intelligentsia \cite{Openlett36:online}, that we consciously do not cover in our scope of harms, and limit our advocacy to a set of present, tangible societal harms that these models pose. We do think there is scope here for future work to address these concerns more directly by taking these dynamics into consideration.

\section{Conclusion}
In this paper, we have discussed the current landscape of generative AI regulation across the world, identified key stakeholders, harms and values that we aim to cover via our framework, and analyzed the advantages and shortcomings of both centralized regulation as well as crowdsourced safety mechanisms. Based on our scoping and priorities, we have defined criteria for a governance framework in the U.S. context that addresses the described shortcomings, and safeguard against the key harms identified. Finally, we have defined a Dual Governance Framework that fulfills the criteria and elaborated on the alliance it creates between centralized regulation and crowdsourced mechanisms. We believe that an implementation of this framework will result in more efficient regulation of generative AI harms, and provide consumers, and small and  big tech companies with equitable capacity to innovate on generative AI, while ensuring equity and safety of the user community.

% \subsection{Future Work and Applicability}

\bibliographystyle{ACM-Reference-Format}
\bibliography{ref}

\appendix

\end{document}